\documentclass[journal]{IEEEtran}

\usepackage{amsmath}
\usepackage{newtxtext}
\usepackage[subscriptcorrection]{newtxmath}
\usepackage{graphicx}
\usepackage{hyperref}

\ifCLASSINFOpdf
\else
\fi

\begin{document}

\title{Ranking of Multi-Response Experiment Treatments}

\author{Miguel R.~Pebes-Trujillo,
	Itamar Shenhar, Aravind Harikumar,\\Ittai Herrmann, Menachem Moshelion, Kee Woei Ng, Matan Gavish
\thanks{M. R. Pebes-Trujillo (E-mail: mpebes@ntu.edu.sg) is with the School of Materials Science and Engineering at Nanyang Technological University, Singapore.}}

\markboth{Preprint}
{Shell \MakeLowercase{\textit{et al.}}: Bare Demo of IEEEtran.cls for IEEE Journals}

\maketitle

\begin{abstract}
We present a probabilistic ranking model to identify the optimal treatment in multiple-response experiments. In contemporary practice, treatments are applied over individuals with the goal of achieving multiple ideal properties on them simultaneously. However, often there are competing properties, and the optimality of one cannot be achieved without compromising the optimality of another. Typically, we still want to know which treatment is the overall best. In our framework, we first formulate overall optimality in terms of treatment ranks. Then we infer the latent ranking that allow us to report treatments from optimal to least optimal, provided ideal desirable properties. We demonstrate through simulations and real data analysis how we can achieve reliability of inferred ranks in practice. We adopt a Bayesian approach and derive an associated Markov Chain Monte Carlo algorithm to fit our model to data. Finally, we discuss the prospects of adoption of our method as a standard tool for experiment evaluation in trials-based research.
\end{abstract}

\begin{IEEEkeywords}
Bayesian modeling, multi-objective optimization, experiment evaluation, paired comparisons.
\end{IEEEkeywords}

\IEEEpeerreviewmaketitle

\section{Introduction}

\subsection{Research Question and Motivation}
\label{sec:research_question}

This article formalizes an answer to the question, {\em Which treatment is the optimal?} The motivating problem arises from agriculture, specifically identifying the best irrigation treatment to achieve desirable crop traits. We outline the context in which this question arises and why answering it is important.

In response to the global food crisis, countries are adopting different strategies to control the vulnerabilities of their food systems. For example, 90\% of Singapore's food supply comes from imports (\cite{Sundram_2023}). To improve its self-reliance, it has set an ambitious goal called the ``30 by 30'' to domestically uplift its ``capability and capacity to sustainably produce 30\% of Singapore's nutritional needs by 2030'' (\cite{SFA_2022}, \cite{SFA_2024}). This goal is inherently challenging because, like most food security policies, it aims to improve multiple targets concurrently. For example, food should be ``sufficient, affordable, and nutritious'' (\cite{Sundram_2023}). However, actually implementing irrigation interventions to produce food that is sufficient, affordable, and nutritious comes with various trade-offs, making the concept of ``optimal policy'' to remain unclear. 

The problem of evaluating interventions from multiple competing criteria, and known ideal responses, is widespread and manifests in various forms. We illustrate this with examples from three fields:

(1) {\em Plant science}. Experiments test which fertilizer maximizes yield and nutrition. Each group of plants receives a different fertilizer during a growth cycle. Then groups are compared using multiple measurements, e.g. dry weight and physiological features such as water and nutrient use efficiency, and chlorophyll levels on the leaves. Data is gathered using plant phenotyping systems including {\em in situ} sensors (e.g. \cite{Halperin_et_al_2017}), remote sensing (e.g. \cite{Harikumar_et_al_2024_A}, \cite{Harikumar_et_al_2024_B}), mineral analysis (e.g. \cite{Shenhar_et_al_2024}), and metadata (e.g. experiment costs). This data reveals the variety of features to optimize, which often lead to competing criteria. Consider the following two-feature illustrations. (a) Suppose we aim to produce leafy crops with maximum weight and leaves that are the greenest possible. However, maximum greenness might come at expense of non-maximum plant weight. (b) We want plants with maximum levels of two minerals, both important for nutrition, but one of them is an inhibitor of the other (e.g. as the patterns observed in \cite{Shenhar_et_al_2024}). (c) Maximum weight might come at the expense of using the costliest fertilizer, resulting in an economic trade-off. In reality, multiple features come into analysis and ranking treatments is highly complex. This problem worsens from sample size constraints caused by the high costs of automatic irrigation, limiting the number of plants per experiment.

(2) {\em Materials science} (MS). New materials are designed to exhibit properties for practical utility. For example, in biomaterials for food packaging ``desirable properties [...] include biodegradability, barrier properties, optimal mechanical strength, resistance to moisture, UV resistance, improved shelf-life, and recyclability'' (\cite{Pratim_et_al_2024}). Typically, a set of variants is produced and the task is to identify the one closest to exhibiting ideal properties, for real use. This is ``challenging as [in MS] the objectives may conflict with each other'' (\cite{Shi_et_al_2023}). Artificial intelligence (AI) and machine learning (ML) methods are gradually permeating multi-property material optimization, though there is still no consensus on a standard solution (\cite{Stergiou_et_al_2023}). Nevertheless, even with such an assistance, the optimal material is chosen by making trade-offs due to inherent property conflicts. Additionally, MS experiments are time- and labor-intensive, and often only a few samples can be afforded.

(3) {\em Reinforcement learning} (RL). The third example is presented as an unexplored application. Consider a RL system with a discrete action space and a continuous, multivariate state space. The relationship between {\em action space} and {\em state space} can be redefined as a {\em treatment} and {\em features} problem. In a multi-armed bandit framework, treatments are selected from a finite set based on their potential to maximize a predefined cumulative reward, which is derived from momentary system features as illustrated by Bellman equations (\cite{Sutton_and_Barto_2018_}). Conversely, under a Thompson sampling approach (\cite{THOMPSON_1933}, \cite{WAHRENBERGER_et_al_1977}, \cite{Chapelle_Li_2011}, \cite{Honda_Takemura_2014}, \cite{Russo_et_al_2018}), the treatment space can be explored with probabilities that depend on their observed performances, jointly encoded as a ranking. This ranking distribution enables stochastic exploration, providing an alternative to heuristic cumulative rewards. We term this principled conceptualization {\em ranking-based RL}. Given its simplicity, we anticipate that it may become prevalent once further explored. To our knowledge, this approach has not yet been reported in the literature.

The described problem appears to be universal, arising in several other areas of application that can be quite distinct. To name a few, fraud detection (e.g. ranking fraudulent cases by multi-source criticality), profiling evaluation in business (e.g. comparing the relative efficiency of different sales channels, or customer segments), and network science (e.g. ranking nodes by a consensus centrality measure).

\subsection{Current practice and literature}
\label{sec:literature_review}

To solve problems like those described in \ref{sec:research_question}, practitioners often rely on pairwise t-tests and analysis of variance (\cite{Wasserstein_Lazar_2016}). This remains a pervasive practice in trials-based fields; even when assumptions, such as population normality with small sample sizes, are rarely satisfied. While these statistical tools do not fully support learning a hierarchy of treatment performances from multiple responses, they are used to compare treatments by property, separately. The choice of an optimal treatment is ultimately shaped by subjective scrutiny of performances across properties.

Although richer methods have been proposed in other arenas, they have never permeated trials-based fields. For example, the problem of eliciting multiple scores for the same set of entities (i.e. resembling the treatments) to produce a consensus score has been studied since the early 1900s. These methods can be categorized into three groups: distributional based, heuristic, and stochastic search (\cite{Lin_2010}). We endorse this classification as both valid and suitable. Heuristic methods use deterministic functions to aggregate individual ranks (e.g. computing aggregated Z-scores). Stochastic search methods optimize a criterion (e.g. a loss function) that commonly encodes how individual contributions are weighted in the aggregation process (e.g. \cite{Cohen_et_al_1998}, \cite{Kujlman_and_Rundensteiner_2020}). Distribution-based methods rely directly or indirectly on the ``law of comparative judgment'', which posits that observing the ordering of two individuals relative to each other sufficiently often enables the estimation of the probability that one prevails over the other (\cite{Thurstone_1927}).

Building rankings by using pairs of individuals as elemental units to analyze competition has been the cornerstone of paired comparison methods. These competitions are intrinsically stochastic as ``better players sometimes lose, and worse players win'' (\cite{Newman_2022}). Since observed outcomes only align with true treatment rankings in terms of overall behavior, addressing this inherent uncertainty requires a statistical solution rather than a fixed heuristic.

Statistical models for paired comparisons trace back to seminal work by \cite{Zermelo_1929} and \cite{Bradley_and_Terry_1952}; which hypothesize that competitors possess inherent {\em strengths}, later referred to as {\em dominance indexes}. These strengths determine the probability of one competitor prevailing over another and subsequently shape the distribution of observable competition outcomes. With no known contemporary parallel, \cite{Davidson_and_Farquhar_1976} provides an extensive reference on the early developments in competitor rankings, including closed-form solutions and hypothesis testing for the strength parameters.

Paired comparisons have also influenced network science, primarily through their Bayesian formulation (e.g.  \cite{Adams_2005},  \cite{Newman_2022}). This is not surprising as networks can serve as sparse representations of competitions. In most cases, not all competitors engage with one another; instead, only a subset of possible competitions actually occurs. These interactions are encoded as a graph, where competitors are represented by nodes and win/loss events are depicted as directed links. From this perspective, a full set of interactions among treatments, i.e. paired comparisons, represents a special case of a fully dense network, where treatments are the nodes and their comparative metrics are shown as weighted, directed links.

Though paired comparison methods have been explored within the discussed frameworks, none has jointly incorporated specific considerations for their application to experiment evaluation. This includes addressing the unique data structure generated by multiple-response experiments, encoding information about ideal outcomes, operationalizing what it means to be the optimal treatment, and providing a cohesive way to report both global and per-response treatment hierarchies. Next, we outline the framework we developed to tackle these specific needs.

\subsection{Overview of our method}
\label{sec:adapting}

\subsubsection{How do we define ``optimal treatment''?}

Our model builds on one fundamental principle: given a set of ideal features, the {\em optimal} treatment is the one that produces outcomes closest to ideal. That means that one treatment is {\em better} than another if it is more prone to generate individuals with the desired properties. Hence, the performance of a treatment depends on a set of evaluation features, and a chosen closeness function (in terms of proximity as defined in \cite{Trosset_2023}, p.12). For multiple evaluation features, the optimal treatment is the one with the best {\em overall} performance. The optimality of a treatment is characterized by its rank or strength: the higher the rank, the better the treatment. We refer to the ordered set of ranks as the {\em treatment ranking}. 

As a note on semantic rigor, \cite{Pebes_2023} recently formalized the distinctions between the mathematical objects related to our interpretation of hierarchies, including rankings. We remark we use the term {\em ranking} in this paper to refer specifically to the order of treatments by (total) strength, rather than in a broader mathematical sense.

\subsubsection{Approach}

We cast our problem as a treatment competition, i.e. treatments are compared pairwise under multiple (possibly conflicting) criteria with the final goal of ranking them. Our model to rank treatments is an adaptation of the Bayesian, hierarchical, and multivariate Bradley--Terry model to work with experiment data. As a Bayesian approach, our method offers guarantees that it asymptotically achieves reliable ranking reconstruction.

The closest and earliest related work is \cite{Davidson_and_Solomon_1972} who used conjugate parameterizations to characterize prior information and developed posterior inference. Subsequent related work includes extracting rankings from sparse unimodal relational data, e.g. \cite{Adams_2005}. Our proposal is also closely comparable to \cite{Newman_2022} and \cite{Newman_2023}, who, using a network science approach, address the problem of inferring global rankings from multivariate criteria.

The parameterization of our model is chosen so that it resembles rank aggregation neatly. To achieve this, we develop a Markov Chain Mont Carlo (MCMC) tailored to fit our model to actual data. Such a parameterization provides an interpretable alternative for reporting the optimality of treatments at global and by-property levels.

Our method is domain-agnostic and generalizable to experiments with (1) feature heterogeneity and, possibly, (2) competing targets and (3) few subjects per treatment, often limiting strong distributional assumptions. We use running examples from plant science and precision agriculture, which not only exhibit these challenges but also hold significant practical importance. While emerging technologies are recognized as key to advancing precision agriculture, sustainability, and food security, interpretable data analysis and non-black-box methods are also essential for supporting regulatory frameworks and ensuring transparency (\cite{hlpe_2022}). Our selection of examples reflects this need.

\subsection{Naming Conventions}
\label{sec:naming_conventions}

Since experiments lie at the center of our research objectives, first we set some basic nomenclature. We run experiments to learn relations between two or more variables and generally one group of variables is assumed to {\em cause} another. The causing ones are commonly referred to as {\em independent variables}, and the caused as {\em responses}. In an experiment most of the aspects are predetermined or set to be fixed, which many times translates as ``controlling the environment''. The single aspect permitted to vary are the independent variables, so that variations in the responses are directly attributed to their casual effects. In this article, every mention of {\em experiment} is referred to those designed (in the sense of \cite{Lawson_2015}) with only one independent variable, and with either one or multiple responses. The different values that the single independent variable can hold are called {\em interventions} or {\em treatments}. In practice, treatments are the different empirical regimes to which the experimental units are subjected to. This includes potential default or {\em control} regimes representing {\em statu quo} or lack of intervention. We use interchangeably the terms {\em experimental units}, {\em subjects}, and {\em individuals}. Finally, we jointly call {\em features} to the set of responses plus any other metadata recorded at individual level. Our work is circumscribed to the definitional considerations described here.

\section{The Treatment Ranking Model} 
\label{sec:model}

\subsection{Data Structure and Heuristics for Defining Ideal Outcomes}
\label{sec:datastructure}

The problem is to determine which of $K$ treatments is more effective in producing a desired set of $M$ features $\boldsymbol{X}^*=\{X^*_1,\dots,X^*_M\} \in  \mathbb{R}^M$. For $k=1,\dots,K$ the $k$th treatment consists of $n_k$ experimental units and hence the vector of sample sizes $\boldsymbol{n}=\{n_1,\dots,n_K\}$ is known. We assume that only a few samples per treatment are available and the members of $\boldsymbol{n}$ can be arbitrarily small. For each individual, we collect $M$ feature measurements. We represent the collected data for all the individuals assigned to the $k$th treatment as a matrix $\boldsymbol{X}_k = [X_{kij}] \in \mathbb{R}^{n_k \times M}$, for fixed $k$, $i=1,\dots,n_k$ and $j=1,\dots,M$.

Each observation $X_{kij}$ describes the value of the $j$th feature of the $i$th individual assigned to the $k$th treatment. Let the centered dot subscript $\cdot$ denote all the indexes of its correspondent position. Then each column vector $\boldsymbol{X}_{k\cdot j} = X_{k1j},\dots,X_{kn_Kj} \in \mathbb{R}^{n_k}$ denotes the measurements of the $j$th feature for all the individuals assigned to treatment $k$. Let $\boldsymbol{X}_{\cdot \cdot j}$ denote the measurements of the $j$th feature for individuals from all treatments, and $\boldsymbol{X}$ denote the sequence of random matrices for all treatments, i.e. $\boldsymbol{X}=\{\boldsymbol{X}_{1},\dots,\boldsymbol{X}_{K}\}$.

We assume the ideal profile $\boldsymbol{X}^*$ is known, and hence {\em deviations} of the observed measurements $\boldsymbol{X}$ from $\boldsymbol{X}^*$ can be computed. Out of many possibilities, we penalize deviations above and below ideal values equally. We call this assumption {\em deviation symmetry}. To operationalize deviation symmetry, we use absolute subtractions of the form $|X^*_j - X_{kij}|$ as our elemental measure of proximity. This might not be suitable when there is a preference towards treatments committing positive or negative deviations. E.g. conditionally defining $D_{kij} = 0$ if $X_j^*<X_{kij}$ and $D_{kij} = X_j^*-X_{kij}$ if $X_j^*>X_{kij}$ dismisses the differences when measurements go above ideal values. Thus, this choice can be customized to accommodate different needs.

Let $D_{kij} = |X^*_j - X_{kij}|$ define deviation random variables. Then we produce the matrix of deviations $\boldsymbol{D}_k = [D_{kij}] \in (\{0\} \cup \mathbb{R}^+)^{n_k\times M}$ whose dimensions equate $\boldsymbol{X}_k$'s.

Analogously to $\boldsymbol{X}$, let $\boldsymbol{D} = \{\boldsymbol{D}_1,\dots,\boldsymbol{D}_K\}$ represent the full data of deviations for all treatments. Each column vector $\boldsymbol{D}_{k\cdot j} = D_{k1j},\dots,D_{kn_Kj} \in (\{0\} \cup \mathbb{R}^+)^{n_k}$ describes the deviations of the $j$th feature for all the individuals assigned to treatment $k$.

In practice we use different heuristics to operationalize desired features $\boldsymbol{X}^*$ to obtain $\boldsymbol{D}$ from $\boldsymbol{X}$. For example, if the interest for certain feature $j$ is to be maximum, we use the highest observed value and define $X^*_j$ as the $n$th order statistic of $\boldsymbol{X}_{\cdot \cdot j}$ where $n=n_1+\cdots+n_k$. That is, we let $X^*_j=\max \{\boldsymbol{X}_{\cdot \cdot j}\}$. Similarly, when the interest for a feature is to be minimum, we use the first order statistic and set $X^*_j=\min \{\boldsymbol{X}_{\cdot \cdot j}\}$. For cases where there is a desirable range $[a,b]$ for a feature measurement to lie in, we define $X^*_j$ as the middle point $(a+b)/2$. We assume this set of rules suffices to represent ideal feature profiles, though certainly other alternatives are admissible.

\subsection{Likelihood}
\label{sec:likelihood}

We determine whether treatments are better or worse by pairwise comparing their deviations $\boldsymbol{D}_1,\dots,\boldsymbol{D}_K$. On average, we expect that better treatments produce smaller deviations and worse treatments larger ones. For a feature $j$, let $r$ and $s$ be two treatments to be compared. Without loss of generality, if $r$ is better than $s$, we expect that values in $\boldsymbol{D}_{r\cdot j}$ are smaller than values in  $\boldsymbol{D}_{s\cdot j}$. For a fixed feature $j$ let $Y_{jrs}$ be the number of times that deviations from $r$ are smaller than deviations from $s$ in within-treatment comparisons. With  $I()$ denoting the indicator function, we define the treatment-comparison variable $Y_{jrs}$ as
\begin{equation}
	Y_{jrs}=\sum_{a=1}^{n_r}\sum_{b=1}^{n_s}I(D_{raj}<D_{sbj}).
	\label{eq:Y}
\end{equation}

Let the three-dimensional array $\boldsymbol{Y} = [Y_{jrs}] \in (\{0\} \cup \mathbb{Z}^+)^{M \times K \times K}$ denote the collection of $Y_{jrs}$ for all features $j$ and all pairs $(r,s)$, and denote by $\boldsymbol{y}$ its observed realizations. In principle, the sampling model for $\boldsymbol{y}$, $p(\boldsymbol{y})$\footnote{We use $p()$ to denote both probability density functions and probability mass functions. The difference is made implicitly via their arguments.} can be any discrete distribution defined on $\{0,1,\dots,n_r \times n_s\}$. In practice, we aim to reasonably relate the true theoretical strengths of treatments to their empirical performances that we express through the observable comparative counts $\boldsymbol{y}$. We define $p(\boldsymbol{y})$ by making two structural assumptions. First, we assume independence over features. Second, for a fixed pair of treatments, we assume that pairwise comparisons of deviations are exchangeable. By the de Finetti theorem (\cite{de_Finetti_1937}, \cite{Diaconis_and_Freedman_1980}, \cite{Bernardo_and_Smith_2000}), we factorize the joint model per feature as the product of observations that are conditionally independent given some parameter vector $\boldsymbol{\theta}$ that suffices for them to be so, and which is random. For $r=1,\dots,K$ and $s=1,\dots,K$ we write the likelihood as
\begin{equation}
    \hspace*{-.7cm}p(\boldsymbol{y} \mid \boldsymbol{n},\boldsymbol{\theta}) = \prod_{j=1}^{M} \prod_{r<s}^K p(y_{jrs} \mid \boldsymbol{n},\boldsymbol{\theta}).
    \label{eq:likelihood}
\end{equation}
In (\ref{eq:likelihood}) the external and internal products come from the first and second structural assumptions, respectively. The internal product performs $\binom{K}{2}$ comparisons. Also, $\boldsymbol{n}$ is known, and we leverage $\boldsymbol{\theta}$ to encode meaningful relations among treatments. The resulting expression for $M=1$ was called the multi-binomial model by \cite{Davidson_and_Solomon_1972}.

To specify $p(y_{jrs} \mid \boldsymbol{n},\boldsymbol{\theta})$ we follow \cite{Bradley_and_Terry_1952} and assume that treatment strengths are represented by positive real values: the higher the value, the better the treatment. In the vast literature on the topic, those values have been called strengths, ratings, weights, or dominance indexes, to name a few. We call them {\em dominance indexes}, as \cite{Adams_2005} does, and reserve the term {\em weights} to decompose those indexes into contributing sub-indexes encoding treatment strengths for each feature. The incorporation of these indexes is discussed next.

\subsection{Dominance Indexes as Parameter Expansion}
\label{dominance_indexes}

The dominance index of a treatment characterizes the propensity of its individuals to exhibit ideal features. Let $d_k \in \mathbb{R}^+$ denote the index for the $k$th treatment, then we assume that the vector $\boldsymbol{d}=d_1,\dots,d_K$ is explanatory for $y_{jrs}$, and encode it as a parameter by letting $\boldsymbol{d} \subset \boldsymbol{\theta}$. We treat the dominance index of each treatment as a global quantity encoding its overall performance, regardless of the number of features $M$.

To conciliate the presence of multiple features with a single global dominance index, we assume each treatment has $M$ associated sub-indexes characterizing its performance under each feature. This has parallels with the analogy in \cite{Bradley_and_Terry_1952} (and other subsequent pieces of work e.g. \cite{Sadasivan_and_Rai_1973}) of multiple judges who might judge one single competition differently; and with the multiple interaction types in \cite{Newman_2022} forming a multiplex-network. Here this translates as comparing treatments under multivariate criteria, i.e. as collecting multiple comparative counts $y_{jrs}$s for the same pair of treatments ($r,s$), for multiple features $j$s.

Though there are many possibilities for defining sub-indexes, we restrict them to being proportions of the global dominance index. That is, for the $k$th treatment, we decompose its dominance index $d_k$ as the sum $d_k=d_{k1}+\hdots+d_{kM}$\footnote{For simplification global and by-feature indexes are both denoted by $d$. The difference is made via the single or double subscript.}. We let each $d_{kj}$ lie in a predefined range $[0,u]$, with lower bound $0$, for convenience, and upper bound $u$. Consequently $d_k$ lies in $(0,u\times M)$. Since we assume sub-indexes are proportions of the global index, then global indexes can be expressed as convex combinations, trivially. Let $\boldsymbol{w}_k=[w_{k1},\dots,w_{kM}]$ be the vector of weights for the $k$th treatment such that $\sum_{j=1}^{M}w_{kj}=1$.  For each feature $j$ we define
\begin{equation}
    d_{kj}=d_k w_{kj}.
    \label{eq:dkj}
\end{equation}

Finally, we let $\boldsymbol{w}=[\boldsymbol{w}_1,\dots,\boldsymbol{w}_K]^\top$ be the $(K\times M)$-matrix encoding the weight vectors for all treatments and assume it is explanatory for $y_{jrs}$, i.e. $\boldsymbol{w} \subset \boldsymbol{\theta}$.

\subsection{The Bradley--Terry Model for Experiment Data}

We have all the ingredients for defining $p(\boldsymbol{y} \mid \boldsymbol{n},\boldsymbol{\theta}=\{d,w\})$. Back to (\ref{eq:likelihood}) and with $Y$ defined as in (\ref{eq:Y}), $Y_{jrs}$ is the sum of Bernoulli distributed random variables, i.e. $Y_{jrs}$ is a Binomial random variable whose probability of success equates to the probability of individual events of the form $D_{rja}<D_{sjb}$. We emphasize that for fixed treatments $(r,s)$ and fixed feature $j$, our structural assumptions set these comparisons to be exchangeable or equivalently independent given the indexes $(d_r,d_s)$ and weights $(w_{rj},w_{sj})$ (or given the sub-indexes $d_{rj}$ and $d_{sj}$ as well).

Using these parameters, we define the probability that a random individual $a$, from treatment $r$, outperforms a random individual $b$, from treatment $s$, at producing closer-to-optimal outcomes for feature $j$. As historically done, we generically call these probabilities {\em prevailing probabilities}. In principle, for two treatments $r$ and $s$, we need any function $\phi:\mathbb{R} \rightarrow (0,1)$ such that it inputs a real-valued function of $d_r$ and $d_s$, and it outputs a probability. In practice, any ``cumulative distribution function of a $\dots$ symmetric random variable'' (\cite{Stern_1990} referring to \cite{David_1988}) does the job. Traditionally, the input function has been set to be the raw difference of the dominance indexes. This choice, though simple, has been rich enough to map pairs of indexes to probabilities and remains the preferred one (from \cite{Bradley_and_Terry_1952} to \cite{Newman_2023}). We do the same and express (\ref{eq:likelihood}) as
\begin{eqnarray}
    \hspace*{-.5cm}
    p(\boldsymbol{y} \mid \boldsymbol{n},\boldsymbol{d},\boldsymbol{w}) &=& \prod_{j=1}^{M} \prod_{r<s}^K {n_r \times n_s \choose y_{jrs}} \nonumber \\
    &\times&\left(\phi\left(d_r w_{rj}-d_s w_{sj}\right)\right)^{y_{jrs}} \nonumber \\ 
    &\times& \left(1-\phi\left(d_r  w_{rj} -d_s w_{sj}\right)\right)^{n_r \times n_s-y_{jrs}}.
	\label{eq:likelihood_full}
\end{eqnarray}

Starting at the internal product in (\ref{eq:likelihood_full}), the expression resembles the Bradley--Terry model with $\phi$ defined as the sigmoid function, i.e. $\phi(x)=(1+e^{-x})^{-1}$. Early roots of this form of likelihood can be seen in the parameterizations revised in \cite{Davidson_and_Solomon_1972}, or \cite{Adams_2005}. The critical difference with our work is the hierarchical definition of global dominance indexes as additive composites of indexes by features that we input into $\phi$. By doing this, we offer a new alternative for treatment strengths information passing in the likelihood:  from the ranking to the observable comparative experiment results $\boldsymbol{y}$, via a prevailing probability that is interpretable at feature-level, as developed in \ref{dominance_indexes}. With the sigmoid assumption, (\ref{eq:likelihood_full}) unfolds as
\begin{eqnarray}
    \hspace*{-.7cm}p(\boldsymbol{y} \mid \boldsymbol{n},\boldsymbol{d},\boldsymbol{w}) &=& \prod_{j=1}^{M} \prod_{r<s}^K {n_r \times n_s \choose y_{jrs}} \nonumber\\
    &\times& \left(\frac{1}{1+e^{-(d_r w_{rj}-d_s w_{sj})}}\right)^{y_{jrs}} \nonumber \\ &\times& \left(\frac{e^{-(d_r w_{rj}-d_s w_{sj})}}{1+e^{-(d_r w_{rj}-d_s w_{sj})}}\right)^{n_r \times n_s-y_{jrs}}.
	\label{eq:lik}
\end{eqnarray}

On the basis of defining the random variables $Y$ as in (\ref{eq:Y}) and weights structure as in \ref{eq:dkj}, expression (\ref{eq:lik}) is what we call the {\em Bradley--Terry Model for Experiments} (BTME). Therefore, provided ideal expected outcomes $\boldsymbol{X}^*$ and real observable outcomes $\boldsymbol{X}$, we say $\boldsymbol{Y}$ is BTME-distributed with parameters $\boldsymbol{d}$ and $\boldsymbol{w}$, and use the shorthand notation
\begin{equation}
	\boldsymbol{Y} \mid \boldsymbol{n},\boldsymbol{d}, \boldsymbol{w} \sim \text{BTME}(\boldsymbol{n}, \boldsymbol{d}, \boldsymbol{w}).
	\label{eq:BTME}
\end{equation}

In section \ref{sec:inference}, we develop the inference method for estimating $\boldsymbol{d}$ and $\boldsymbol{w}$ from $\boldsymbol{y}$ using (\ref{eq:lik}). These parameters jointly carry the following information: (1) global ranking of treatments (via $\boldsymbol{d}$), (2) ranking of treatments by feature (via $\boldsymbol{d}$ and $\boldsymbol{w}$), and (3) contribution of each feature to global treatment rankings (via $\boldsymbol{w}$). Identifying the best treatment is a direct consequence of (1).

\section{MCMC Estimation}
\label{sec:inference}

\subsection{Model Parameterization}

We choose to parameterize the BTME generative process and state it constructively as:
\begin{eqnarray}
    d_1,\dots,d_K &\stackrel{iid}{\sim}& p(d) \quad (k=1,\dots,K); \label{step_1}\\
    w_{k1},\dots,w_{kM} &\stackrel{}{\sim}& \textrm{Dirichlet}(\alpha_1,\dots,\alpha_K) \quad \label{step_2} \\
    && (\alpha_1,\dots,\alpha_K \in {\mathbb{R}^+}^K); \nonumber\\
    \pi_{rs} &=& \textrm{Sigmoid}(d_r w_{rj}-d_s w_{sj}) \quad  \label{step_3} \\
    &&(j=1,\dots,M; r < s);\nonumber\\
    y_{jrs} \mid n_r, n_s, \pi_{rs}  &\stackrel{ind}{\sim}& \text{Binomial}(n_r\times n_s,\pi_{rs}) \quad \label{step_4} \\
    && (j=1,\dots,M; r < s); \nonumber
\end{eqnarray}

where $w_{k1},\dots,w_{kM}$ is in the $(M-1)$\textrm{-simplex} for all $k$, and the choice of a Dirichlet distribution provides conjugacy. The Binomial distribution has a standard parameterization. Steps (\ref{step_3})-(\ref{step_4}) represent the unfolded version of the BTME in (\ref{eq:BTME}), or the distribution of $\boldsymbol{Y}$ given $\boldsymbol{d}$ and $\boldsymbol{w}$, in (\ref{eq:lik}), stated hierarchically. The common distribution $p(d)$ is defined on $[0,u \times M]$. We let this to be Uniform$(0,u \times M)$, which does not inform about any prior believes for the treatment performances. This choice is a necessity in our setup characterized by small sample sizes, as otherwise the effect of an informative prior distribution will dominate the effect of data in the posterior inference. The $iid$ assumption for the common distribution in (\ref{step_1}) is not really a requirement but a convenient choice. If needed, dependencies among ranks motivated by domain understanding can be incorporated by jointly modeling $p(\boldsymbol{d})$, and deriving a reasonable factorization for it.

\subsection{\label{sec:estimators}Ranks and Weights Estimators}

We use the joint model $p\left(\boldsymbol{y},\boldsymbol{n},\boldsymbol{d},\boldsymbol{w}\right)$ under a Bayesian approach. We condition the unseen parameters on the observed data to obtain the joint posterior distribution of dominance indexes and feature weights, $p\left(\boldsymbol{d},\boldsymbol{w} \mid \boldsymbol{y},\boldsymbol{n} \right)$, via the proportionality relation
\begin{equation}
	\hspace*{-.7cm}p(\boldsymbol{d}, \boldsymbol{w} \mid \boldsymbol{y}, \boldsymbol{n}) \propto p(\boldsymbol{y}\mid \boldsymbol{n},\boldsymbol{d},\boldsymbol{w}) p(\boldsymbol{d})p(\boldsymbol{w}).
	\label{eq:posterior_prop}
\end{equation}
In (\ref{eq:posterior_prop}), $p(\boldsymbol{y}\mid \boldsymbol{n},\boldsymbol{d},\boldsymbol{w})$ is given by the BTME likelihood in (\ref{eq:lik}). The joint prior $p(\boldsymbol{d},\boldsymbol{w})$ is written as $p(\boldsymbol{d})p(\boldsymbol{w})$ as we assume $\boldsymbol{d}$ and $\boldsymbol{w}$ are a-priori independent.

For the given priors $p(\boldsymbol{d})$ and $p(\boldsymbol{w})$ we compute posterior estimates for the dominance indexes $\hat{\boldsymbol{d}} =\{\hat{d}_1,\dots,\hat{d}_K\}$. We define each entry as the expectation of its correspondent posterior marginal distribution, i.e.
\begin{eqnarray}
\{\hat{d}_1,\dots,\hat{d}_K\} &=&\{E \left(d_1 \mid \boldsymbol{y}, \boldsymbol{n} \right),\dots,\text{  }E \left(d_K \mid \boldsymbol{y}, \boldsymbol{n} \right)\}, \quad \nonumber \\
 &&\text{subject to: } \min(d_1,\dots,d_K)=0.
 \label{eq:est}
\end{eqnarray}
We constrain the minimum index to be equal to 0 to ensure the model is identifiable. Each posterior marginal $p(d_k \mid \boldsymbol{y}, \boldsymbol{n})$ for $k=1,\dots,K$ is obtained after integrating out the joint posterior over all other dominance indexes and all weights, i.e.
\begin{eqnarray}
p(d_k \mid \boldsymbol{y}, \boldsymbol{n}) &=& \int_{d_K} \cdots \int_{d_{k+1}}\int_{d_{k-1}} \cdots \int_{d_1}  \int_{\tilde{w}} p(\boldsymbol{d},\boldsymbol{w} \mid \boldsymbol{y}, \boldsymbol{n}) \nonumber\\
	&& \text{d}\tilde{w} \text{d} d_1 \dots \text{d}d_{k-1} \text{d}d_{k+1}\dots \text{d}d_K.
	\label{eq:estimates}
\end{eqnarray}
The integration over $\tilde{w}$ is a shortcut notation for the integration over every entry of $w_k$ and every $k$. The integrals for dominance indexes go over the interval $[0,u \times M]$ which bounds are constructed so that the prevailing probabilities lie in $(0,1)$ reasonably\footnote{In practice, we are parsimonious and pick the bounds $l$ and $u$ to reach a third digit approximation to it, i.e. $(0.001,0.999)$}.

For $k=1,\dots,K$, we define the estimates for the weights $\hat{\boldsymbol{w}}_k = \{\hat{w}_{k1},\dots,\hat{w}_{kM}\}$ also as the expectations of their correspondent posterior marginal distributions, i.e.
\begin{equation}
	\{\hat{w}_{k1},\dots,\hat{w}_{kM}\} = \{E \left(w_{k1} \mid \boldsymbol{y}, \boldsymbol{n} \right),\dots,E \left(w_{kM} \mid \boldsymbol{y}, \boldsymbol{n} \right)\}.
	\label{eq:est_w}
\end{equation}

Each posterior marginal $p(w_{kj} \mid \boldsymbol{y}, \boldsymbol{n})$ for $j=1,\dots,M$ is obtained after integrating out the joint posterior over all other weights and all dominance indexes, i.e.
\begin{eqnarray}
 p(w_{kj} \mid \boldsymbol{y}, \boldsymbol{n}) &=& \int_{\tilde{w}_{K}} \cdots \int_{\tilde{w}_{k+1}} \int_{w_{kM}} \cdots \int_{w_{k,j+1}} \int_{w_{k,j-1}} \nonumber \\ 
 &&\cdots \int_{w_{k1}} \int_{\tilde{w}_{k-1}} \cdots \int_{\tilde{w}_1}  \int_{\tilde{d}} p(\boldsymbol{d},\boldsymbol{w} \mid \boldsymbol{y}, \boldsymbol{n}) \nonumber \\ \nonumber
	&&  \text{d}\tilde{d} \text{d} \tilde{w}_1 \dots \text{d}\tilde{w}_{k-1} \text{d}w_{k1} \dots \\ \nonumber && \text{d}w_{k,j-1} \text{d}w_{k,j+1} \dots \\&&\text{d}w_{kM} \text{d}\tilde{w}_{k+1}\dots \text{d}\tilde{w}_K.
	\label{eq:estimates_w}
\end{eqnarray}
The integration over $\tilde{d}$ is a shortcut notation for the integration over every entry of $\boldsymbol{d}$. Similarly, integrations over $\tilde{w}_k$s go over every entry of $\boldsymbol{w}_k$.

The produced estimates $\hat{\boldsymbol{d}}$ and $\hat{\boldsymbol{w}}=\{\hat{\boldsymbol{w}}_1,\dots,\hat{\boldsymbol{w}}_K\}$ solve the task of obtaining a sequence of dominance indexes and weights that are the most consistent with both the observed experiment results, and the set of ideal expected outcomes. Our research question on finding the optimal treatment is answered through reporting the maximum a posteriori (MAP) estimate of the distribution on the space of total (descendant) orders defined as $\left \{\{k:d_k=d_{(K)}\},\dots,\{k:d_k=d_{(1)}\} \right \}$ where $d_{k}$ represents the $k$th ordered statistic of the sequence $\boldsymbol{d}$. Let $\boldsymbol{o}$ be a particular total order of the described form, i.e. $\boldsymbol{o} \in$ $\{$space of permutations of $\{1,\dots,K\} \}$. Then for an arbitrarily large number of samples $S$ (which will be formally defined in \ref{sec:posterior_approximation}) from such a posterior distribution, its estimated (posterior) probability is 
\begin{eqnarray}
&&\hspace*{-1.1cm}pr(\boldsymbol{o}) \approx (1/S) \times \nonumber \\ 
&&\hspace*{-1.1cm}\sum_{i=1}^{S} I \left( \left\{\{k:d^{(i)}_k=d^{(i)}_{(K)}\},\dots,\{k:d^{(i)}_k=d^{(i)}_{(1)}\} \right\} = \boldsymbol{o} \right).
\end{eqnarray}
The mode of this distribution (the MAP) provides the estimated final order of treatments by efficacy, where the first element is the inferred best treatment. This choice of declaring the best is more a heuristic choice than a strict derived criterion. E.g. another reasonable alternative is to pick the best treatment based on their posterior marginal probability.

\subsection{Joint Posterior Approximation Algorithm}
\label{sec:posterior_approximation}

To compute estimates for (\ref{eq:estimates}) and (\ref{eq:estimates_w}) we derive a Metropolis-Hastings (M-H) algorithm (\cite{Metropolis_et_al_1953}, \cite{Hastings_1970}) to approximate $p(\boldsymbol{d},\boldsymbol{w} \mid \boldsymbol{y}, \boldsymbol{n})$ as, evidently, no closed-form solutions are available for it.

First, we describe the estimation of dominance indexes. Let $S\in \mathbb{Z}^+$ be an arbitrarily large and let $\boldsymbol{d}^{(1)}=d_1^{(1)},\dots,d_K^{(1)}$ be a vector of initial guesses for the ranks, set deterministically. To produce $\boldsymbol{d}^{(2)}\dots,\boldsymbol{d}^{(S)}$ we iterate as follows. Let $s=2,3,\dots$; for every new proposed set of ranks $\boldsymbol{d}^*=d_1^*,\dots,d_K^*$ drawn from a proposal distribution $q(\boldsymbol{d}^* \mid \boldsymbol{d}^{(s-1)})$, we accept $\boldsymbol{d}^*$ as a new sample, i.e. we assign $\boldsymbol{d}^{(s)} := \boldsymbol{d}^*$, with M-H probability
\begin{equation}
	\min\left\{1,\frac{p\left(\boldsymbol{y} \mid  \boldsymbol{n},\boldsymbol{d}^*,\boldsymbol{w}^{(s-1)}\right)p\left(\boldsymbol{d}^*\right)}{p\left(\boldsymbol{y} \mid \boldsymbol{n},\boldsymbol{d}^{(s-1)},\boldsymbol{w}^{(s-1)}\right)p\left(\boldsymbol{d}^{(s-1)}\right)} \frac{q\left(\boldsymbol{d}^{(s-1)} \mid \boldsymbol{d}^*\right)}{q\left(\boldsymbol{d}^* \mid \boldsymbol{d}^{(s-1)}\right)} \right\}.
	\label{eq:accep_prob_d}
\end{equation}
The terms of the form ``likelihood $p\left(\boldsymbol{y} \mid \boldsymbol{n},\boldsymbol{d},\boldsymbol{w}\right)$ times rank prior $p\left(\boldsymbol{d}\right)$'' in (\ref{eq:accep_prob_d}) come from the proportionality of posteriors in (\ref{eq:posterior_prop}). The proposal $q(\boldsymbol{d}^* \mid \boldsymbol{d}^{(s-1)})$ updates one entry at a time. Step \ref{eq:accep_prob_d} is executed sequentially for each treatment index $k = 1,\dots,K$ of $\boldsymbol{d}$. For each $d_k^*$, we let the proposal $q(d_k^* \mid d_k^{(s-1)})$ to be a truncated normal distribution on $(0,u\times M)$ with mean $d_k^{(s-1)}$, and standard deviation that we fix for convenience. The counterpart term $q(d_k^{(s-1)} \mid d_k^*)$ is also defined as a truncated normal distribution, with mean $d_k^*$ with the same standard deviation. The constraint in (\ref{eq:est}) is implemented by setting  $\boldsymbol{d}^*:=\left(\boldsymbol{d}^*-\min(\boldsymbol{d}^*) \right)$ right before the acceptance/rejection step.

Next, we describe the estimation of weights. For each $k$, let $\boldsymbol{w}_k^{(1)}=w_{k1}^{(1)},\dots,w_{kM}^{(1)}$ be a vector of initial guesses for the ranks, set deterministically. To produce $\boldsymbol{w}_k^{(2)}\dots,\boldsymbol{w}_k^{(S)}$ we iterate as follows. Let $s=2,3,\dots$; for every new proposed set of weights $\boldsymbol{w}_k^*=w_{k1}^*,\dots,w_{kM}^*$ drawn from a joint proposal distribution $q(\boldsymbol{w}_k^* \mid \boldsymbol{w}_k^{(s-1)})$, we accept $\boldsymbol{w}_k^*$ as a new sample, i.e. we do $\boldsymbol{w}_k^{(s)} := \boldsymbol{w}_k^*$, with M-H probability
\begin{equation}
	\min\left\{1,\frac{p\left(\boldsymbol{y} \mid  \boldsymbol{n},\boldsymbol{d}^{(s)},\boldsymbol{w}_k^*\right)p\left(\boldsymbol{w}_k^*\right)}{p\left(\boldsymbol{y} \mid \boldsymbol{n},\boldsymbol{d}^{(s)},\boldsymbol{w}_k^{(s)}\right)p\left(\boldsymbol{w}_k^{(s)}\right)} \frac{q\left(\boldsymbol{w}_k^{(s)} \mid \boldsymbol{w}_k^*\right)}{q\left(\boldsymbol{w}_k^* \mid \boldsymbol{w}_k^{(s)}\right)} \right\}.
\end{equation}
We specify $q(\boldsymbol{w}_k^* \mid \boldsymbol{w}_k^{(s-1)})$ as a random walk in the $(M-1)$-simplex using the prior Dirichlet distribution in \ref{step_2}.

We iterate the algorithm for an arbitrarily large number of steps $S$ that makes the process to achieve the stationary distribution of the posterior. We collect $S$ duple samples from  $p(\boldsymbol{d},\boldsymbol{w} \mid \boldsymbol{y}, \boldsymbol{n})$, and compute estimates for both parameters. We produce a point estimate for (\ref{eq:est}) by setting; for $k=1,\dots,K$; $\hat{d}_k :=\left(\sum_{i=1}^{S}d^{(i)}_k\right)/S$.  Similarly, we produce a point estimate for 	(\ref{eq:est_w}) by setting; for each $k$ and $j = 1,\dots,M$;  $\hat{w}_{kj} :=\left(\sum_{i=1}^{S}w^{(i)}_{kj}\right)/S$. These point estimators are consistent and are guaranteed to converge to the true underlying treatment ranks and weights, respectively, due to the weak law of large numbers. Applied to our case, it states that as $S$ tends to infinity the averages of rank samples will approximate their true expected value, i.e. $\frac{1}{S} \sum_{i=1}^{S}d^{(i)}_k \stackrel{p}{\rightarrow} E(d_k) = \int_{l}^{u} d_k p(d_k) \text{d}d_k$, where $p(d_k)$ represents the marginal posterior distribution of the $k$th treatment dominance index. Similar guarantees apply for the weight estimates.

\section{Simulation Study}
\label{sec:simulation}

\subsection{Experiment Setup}

We demonstrate the effectiveness of our method by designing and executing a prediction test. We generate random data representing comparisons of treatment performances (as defined in (\ref{eq:likelihood}) from known rankings and weights. Then we show how the reconstructed rankings and weights closely approximate the originals. We also discuss some guarantees about our method's performance under scenarios involving an increasing number of treatments and features. 

We simulate artificial data from the theoretical generative process described in steps from (\ref{step_1}) to (\ref{step_4}). For empirical utility, the scale of the simulation mimics the scale of real plant phenotyping experiments (as the ones treated in section \ref{sec:application}). We vary the number of treatments and let $K$ take the values $2,4,6,8,$ and 10. We set the number of experimental units per treatment to ten, i.e. $n_1=10,n_2=10,\dots,n_K=10$. This setup induces up to $\binom{10}{2}$ treatment comparisons and, for each pair of treatments, a hundred comparisons of individuals are performed (to produces $Y$s, as in (\ref{eq:Y})). We vary the number of features and let $M$ take the values $1,2,3,5,$ and $10$. The goal is to check how the accuracy of the ranking reconstruction is preserved as the number of treatments and responses increase. The feature dominance indexes are set to range between 0 and 10, thus the global index varies from 0 to $10\times M$.

The assessment of inference quality is as follows. We run one million MCMC samples to estimate the latent parameters on each of the described scenarios. We assume these executions suffice as a proof-of-concept for the algorithm to reach the stationary distribution of the joint posterior (no burning period or thinning strategy were used). For each scenario, the inferred parameters are the dominance indexes (a vector of length $K$) and the weights (a row-stochastic $(K \times M)$-matrix). For each scenario and for both parameters, rankings and weights, we compute estimation errors as the raw subtraction between true values and estimates. Then we display the distribution of the errors by treatment for comparison. Ultimately, we report the Spearman correlations (defined in $[-1,1]$) between two vectors: the theoretical and the inferred dominance indexes. Here a correlation is high if the inferred treatment dominance index order is mostly consistent with the one for the theoretical one.

\begin{figure*}[!t]
\centering
\includegraphics[width=.9\textwidth]{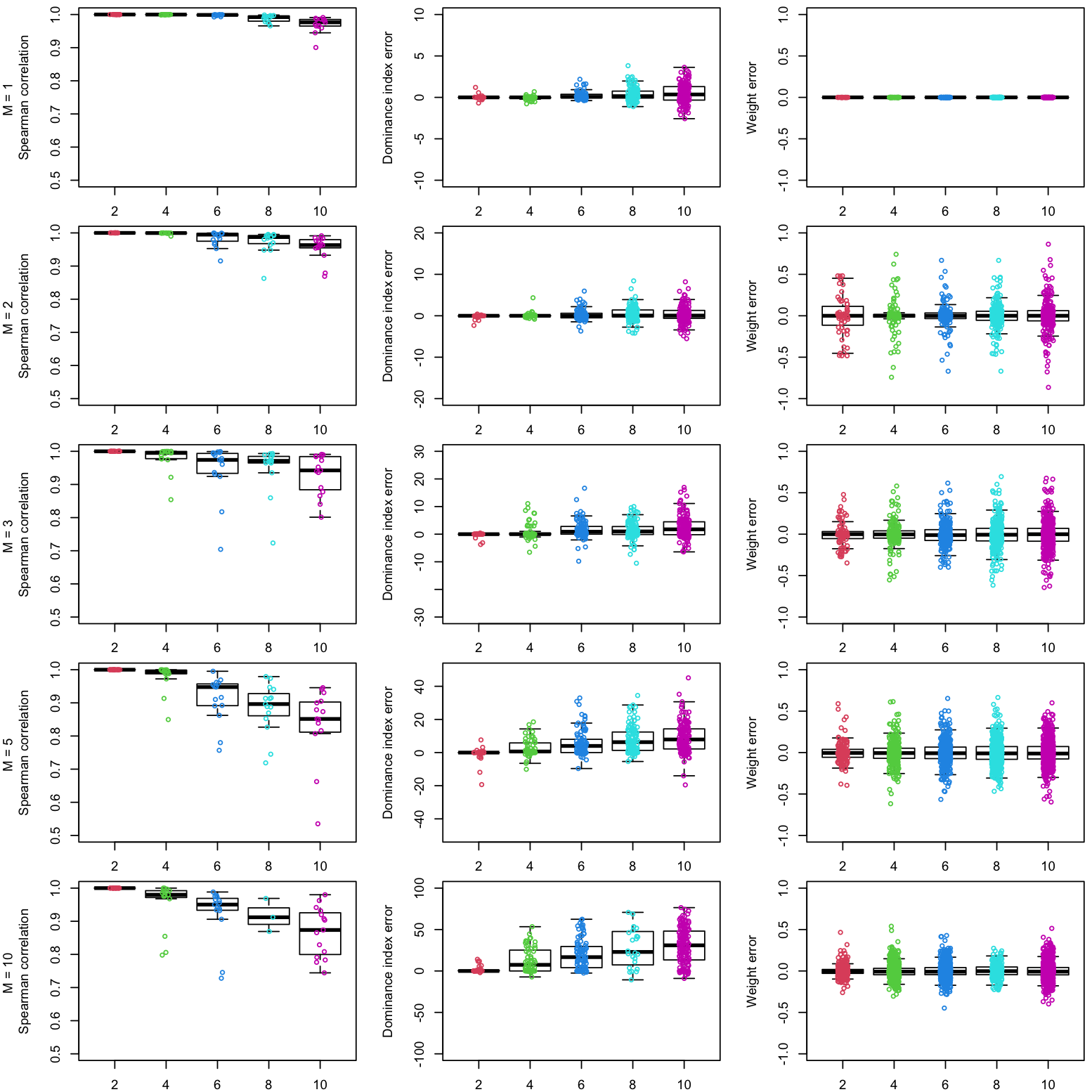}\\
{\fontsize{7pt}{10pt}\selectfont Number of treatments K}
\caption{Reconstruction of rankings and weights. Rows vary over number of features $M=1,2,3,5,10$. The x-axis, varies over number of treatments $K=2,4,6,8,10$. Column 1: Spearman correlations of true versus reconstructed rankings. Column 2: Distribution of differences between true and inferred dominance indexes. Column 3: Distribution of differences between true and inferred weights.}
\label{fig:reconstruction}
\end{figure*}

\subsection{Predictive Results}

Figure \ref{fig:reconstruction} shows the reconstruction results organized in three columns, and each graph shows the algorithm performance trend as we increase the number of treatments. Each row corresponds to a different number of features, $M$, under consideration.

The first column shows the distribution of Spearman correlations, where we set the y-axis to start at 0.5. Results indicate that our method is effective in inferring true latent rankings from data. This is supported by observed correlations which, apart from a few outliers, mostly lie above 0.7 over the whole spectrum of our experiments. 

The second column shows the distribution of the estimation error for the dominance indexes shown in its natural scale. The third column shows the distribution of the estimation error for the feature weights. Our results show that the error committed when reconstructing dominance indexes is small. Errors are centered at zero and the interquartile range (IQR) is small at each correspondent scale. We also observe that achieving accurate reconstruction becomes more difficult as we increase the number of treatments, which also correlates with increasing variability. The reconstruction of weights is less sensitive. The error variability increases only slightly, and it is mostly preserved when increasing the number of treatments. This indicates that our additive model that defines global indexes in terms of indexes by-feature suffices to represent reliably the decomposition of the strength over features. Finally, dominance index errors in column 2 show no bias of our method on overrating or underrating treatments as values are centered around zero. Analogously, weight errors in column 3 show no bias of our method to overweight or underweight features.

From the graph we can also see reflected the computational time complexity carried by our model implementation. Focused on the figures displaying dominance index correlations, we see a slight performance degradation while increasing the number of treatments, going horizontally from left to right in the first figure. At the same time, we see almost no performance degradation when looking at the first boxplot of each graph from the first column (the one for two treatments). This is expected as the time complexity for the log-likelihood function (correspondent to (\ref{eq:lik}) is $O(M \times K^2)$, i.e. it is linear in number of features and quadratic on the number of treatments. Therefore, adding a treatment is computationally more costly than adding a feature into the assessment, and the performance of the inference directly depends on this consideration.

\begin{figure*}[t]
	\begin{center}
		\begin{tabular}{ccc}
			\includegraphics[height=.3\textwidth]{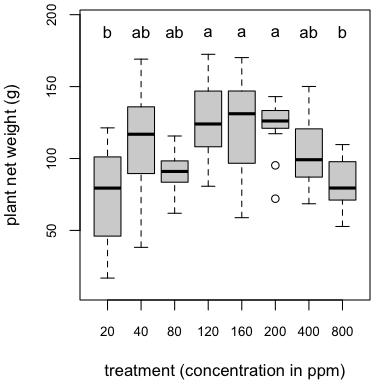} &
			\includegraphics[height=.3\textwidth]{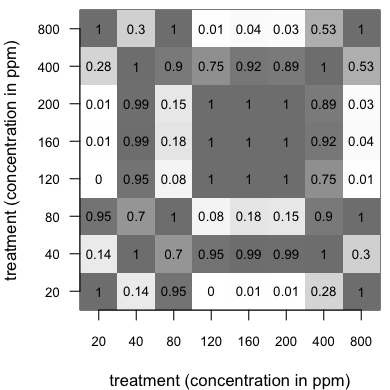} &
			\includegraphics[height=.3\textwidth]{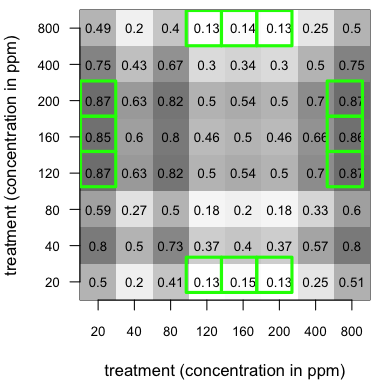}
		\end{tabular}
	\end{center}
	\vspace{-.2cm}
	\caption{Plant net weight analysis: correspondence between Tukey test p-values and the posterior prevailing probabilities of producing plants with higher net weight. Left: distribution of plant net weight by treatment and Tukey test. Center: pairwise p-values for the Tukey test. Right: prevailing probabilities where the ones correspondent to significant entries have been highlighted. Notice how prevailing probabilities as high as 0.82 are not considered significant under the classical approach. Thus, under small sample sizes the BTME produces a more truthful quantitative basis for decision making.}
	\label{fig:netweight_results}
\end{figure*}

\begin{figure*}[]
	\begin{center}
		\begin{tabular}{c}
			\includegraphics[width=.85\textwidth]{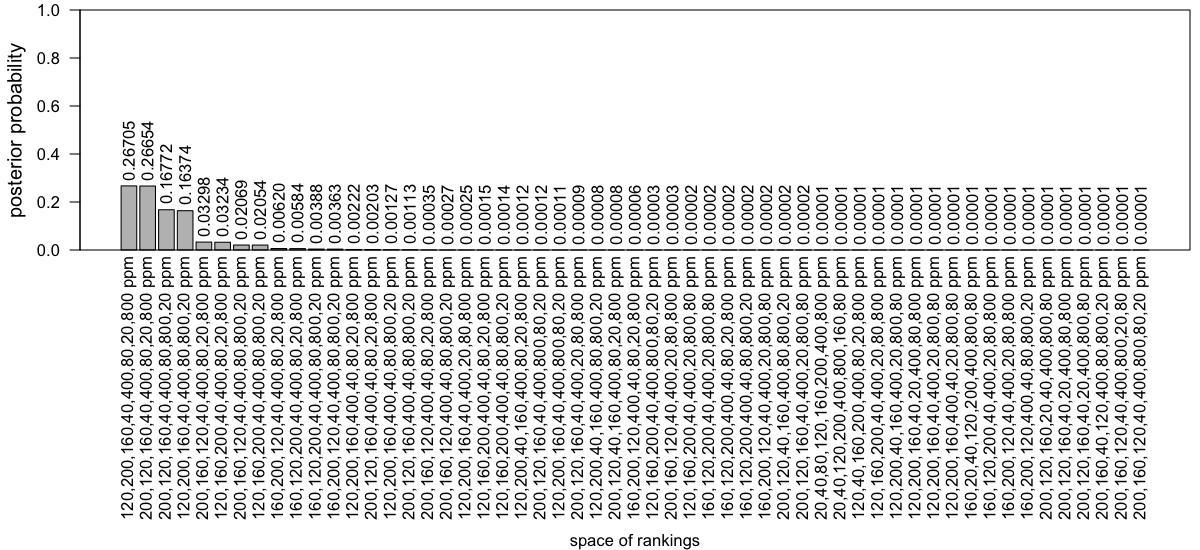}
		\end{tabular}
	\end{center}
	\vspace{-.2cm}
	\caption{Posterior distribution of ranking by maximum plant net weight in grams. The best treatments are 120 and 200 ppm.}
	\label{fig:netweight_barplot}
\end{figure*}

In summary, our simulations quantitatively demonstrate the reliable reconstruction of latent treatment rankings using our method. This reliability is contingent on data whose scale is at least equivalent to that of the simulations. Being backed up by these results, in section \ref{sec:application} we apply our method to applications where no quantitative ground truth is available, but validation comes from domain expertise.

\section{Application on Plant Phenotyping Data}
\label{sec:application}

\subsection{Univariate Data}
\label{sec:univariate}

We conducted an experiment to learn the optimal concentration of a dissolved fertilizer so that plants exhibit desirable traits. The chosen plant and fertilizer are {\em round bayam} (spinach) and nitrogen, respectively. The experiment consists of eight irrigation treatments. Each contains nitrogen at a different concentration: 20 parts per million (ppm), 40 ppm, 80 ppm, 120 ppm, 160 ppm, 200 ppm, 400 ppm, and 800 ppm. For a comprehensive description of the experiment setup see \cite{Shenhar_et_al_2024} and \cite{Harikumar_et_al_2024_A}.

Consider the problem of ranking those treatments according to a single response collected after harvesting. Three features are analyzed separately: {\em yield}, concentration of {\em calcium} on leaves, and concentration of {\em nitrogen} on leaves. To gain intuition about our method's outcomes, each of these traits is analyzed from a practitioner point of view and the conclusions are contrasted with the inferred prevailing probabilities and ranking.

\begin{figure*}[!t]
	\begin{center}
		\begin{tabular}{ccc}
			\includegraphics[height=.3\textwidth]{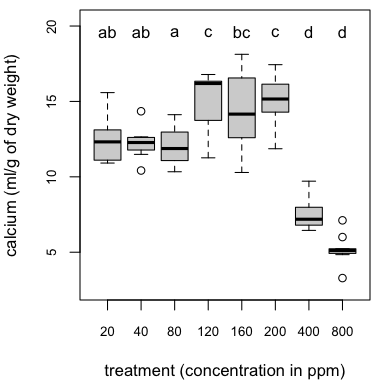} &
			\includegraphics[height=.3\textwidth]{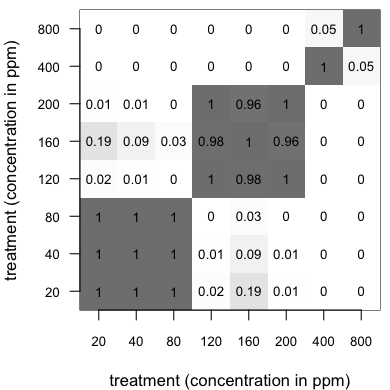} &
			\includegraphics[height=.3\textwidth]{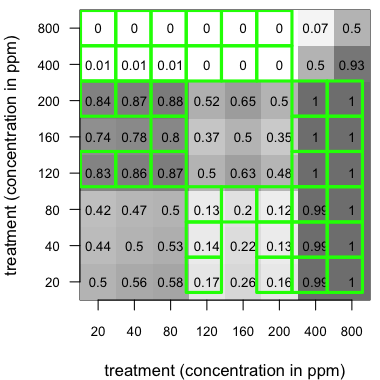}
		\end{tabular}
	\end{center}
	\vspace{-.2cm}
	\caption{Calcium concentration analysis: correspondence between Tukey test p-values and the posterior prevailing probabilities of producing plants with higher calcium concentration in their leaves. Left: distribution of plant net weight by treatment and Tukey test. Center: pairwise p-values for the Tukey test. Right: prevailing probabilities where the ones correspondent to significant entries have been highlighted.}
	\label{fig:calcium_results}
\end{figure*}

\label{sec:netweight}
\subsubsection{Plant net weight analysis}

First, we analyze yield expressed as plant net weight in grams\footnote{From the point of view of plant physiology, a preferable measurement for yield is plant biomass expressed as dry weight. However, for pedagogic purposes, net weight data offers a more interesting optimality pattern to analyze.}. The evaluation criterion is that we want it to be maximum. Domain knowledge suggests that the range spanned by the chosen treatments above is broad and includes concentrations that represent deficit and excess (e.g. as described in \cite{Taiz_et_al_2015} and  \cite{Singh_and_Sharma_2014}). That is, an optimal concentration level is expected to be found in more-central concentrations. Figures \ref{fig:netweight_results} and \ref{fig:netweight_barplot} show our results for yield data.

The leftmost figure in \ref{fig:netweight_results} shows that treatments ranging from 40 ppm to 400 ppm are not statistically different according to a classic Tukey test at a significance level of 0.05. Both the deficit treatments and the excess treatments, i.e. 20 ppm, 40 ppm, 80 ppm, 400 ppm, and 800 ppm, are also not significantly different from each other and thus they lead to the same observable yield patterns. Therefore, if we need yield to be maximum, then reporting any treatment in $\{$40, 80, 120, 160, 200, 400 ppm$\}$ as the optimal would suffice in an assessment based solely on net weight. The middle figure in \ref{fig:netweight_results} shows the pairwise p-values for testing the hypotheses of the form $(H_0:\mu_r = \mu_s, H_1:\mu_r \neq \mu_s)$, for any two treatments $r$ and $s$ (the matrix indexes for each entry) with theoretical means $\mu_r$ and $\mu_s$, respectively.

The rightmost figure in \ref{fig:netweight_results} shows the (directional) comparative probabilities for treatments. Both axes display the treatment values. The value of each entry corresponds to the probability that the row-treatment outperforms the column-treatment. For example, the entry for row 200 ppm and column 20 ppm is 0.87. This is interpreted as ``the probability that a random individual from treatment 200 ppm outperforms a random individual from treatment 20 ppm is 0.87'', which happens to be high. The color indicates how strong or weak these probabilities are. Intuitively, the center of the matrix displays darker row lanes because middle range treatments are better at producing maximum net weight. All the entries in the diagonal display the value 0.5 as the probability that a random individual of certain treatment outperforms another random individual of the same treatment is (just as tossing a fair coin) 0.5. Entries whose correspondent entry in the p-values matrix is less or equal to 0.05 have been highlighted.

Figure \ref{fig:netweight_barplot} shows the posterior distribution of rankings displaying in the x-axis the set of rankings most supported by the observed data, in descendant order by their corresponding posterior probability. Our method reports as the most likely rankings (the MAP estimate) to (120, 200, 160, 40, 400, 80, 20, 800 ppm) and (200, 120, 160, 40, 400, 80, 20, 800 ppm) with probabilities that are indistinguishable for empirical purposes, both $\approx 0.26$. Another relevant observation is that the 160-ppm treatment ranks consistently in the third position in the top four rankings. Though such treatment has the highest median it also has the highest variance among the 120, 160, and 200-ppm treatments and thus is not favored over 120 and 200 ppm. Thus, overall, our method has allocated more probability in more-central treatments, as expected. Going from left to right, the differences between the rankings and the MAP ranking systematically increase, as their posterior probabilities decrease. Clearly the x-axis does not exhaust the space of rankings and those with negligible probability are not shown.

\subsubsection{Calcium analysis}

Mirroring the first example, figure \ref{fig:calcium_results} shows the analysis of calcium concentration on the leaves measured in milligrams per gram of dry weight. The evaluation criterion is that we want it to be maximum as it is directly related to nutritional value. Following the same logic, the leftmost figure shows that maximum values for calcium on leaves are achieved by the treatments 120, 160, and 200 ppm, which are not statistically significant from one another. The middle figure shows the p-values supporting the preceding statement. Comparative probabilities in the rightmost figure shows that treatments 400 and 800 ppm are easily beaten by any of the other treatments and the former is superior to the latter.

\begin{figure}[h]
	\begin{center}
		\begin{tabular}{c}
			\includegraphics[height=.38\textwidth]{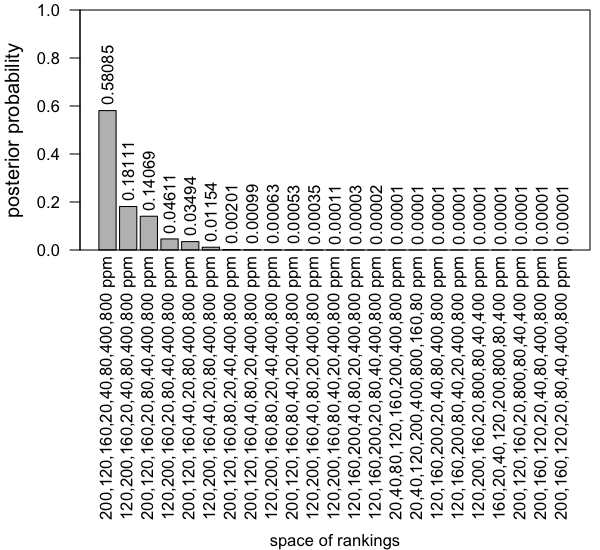}
		\end{tabular}
	\end{center}
	\vspace{-.2cm}
	\caption{Posterior distribution of rankings by maximum calcium concentration.}
	\label{fig:calcium_barplot}
\end{figure}

Figure \ref{fig:calcium_barplot} shows the posterior distribution of rankings by calcium levels. The MAP estimate is (200, 120, 160, 40, 20, 80, 400, 800 ppm). Again, as shown, results are consistent with what visual inspection suggests and with what a classic analysis dictates. 

\begin{figure*}[!t]
	\begin{center}
		\begin{tabular}{ccc}
			\includegraphics[height=.3\textwidth]{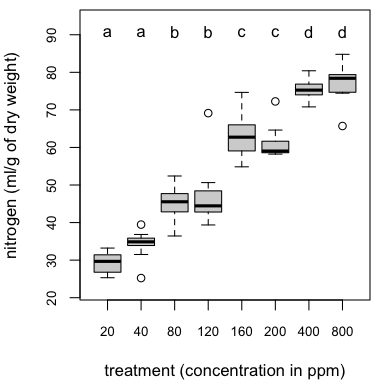} &
			\includegraphics[height=.3\textwidth]{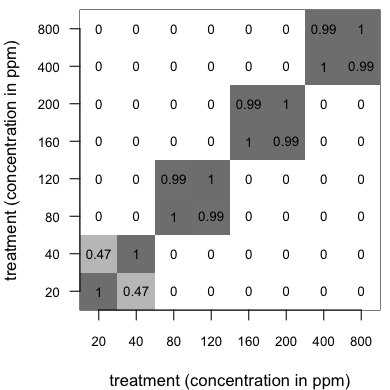} &
			\includegraphics[height=.3\textwidth]{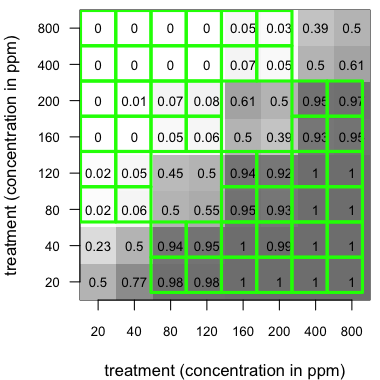}
		\end{tabular}
	\end{center}
	\vspace{-.2cm}
	\caption{Nitrogen concentration analysis: correspondence between Tukey test p-values and the posterior prevailing probabilities of producing plants with lower nitrogen concentration in their leaves. Left: distribution of nitrogen concentration by treatment and Tukey test. Center: pairwise p-values for the Tukey test. Right: prevailing probabilities where the ones correspondent to significant entries have been highlighted.}
	\label{fig:posterior_matrix3}
\end{figure*}

\subsubsection{Nitrogen analysis}

The third example is the analysis of nitrogen concentration on the leaves measured in milligrams per gram of dry weight. In this case, we want nitrogen to be minimum to exemplify the real problem of a feature with potential negative effects. E.g. nitrogen-based fertilizers used huge amounts of energy to be produced and have significant implications in the environment (e.g. see \cite{Wegahita_et_al_2020}). The distribution per treatment in the leftmost figure shows a relatively increasing pattern of nitrogen concentration on leaves as we increase the concentration of the treatment. Naturally, if we target minimizing the nitrogen concentration on leaves without taking into consideration any other feature, then we can expect that the 20-ppm treatment is favored by the algorithm, and that higher concentrations on the treatment are the least favored. The MAP estimate reported in the middle figure is 20, 40, 80, 120, 200, 160, 400, 800 ppm with a posterior probability of 0.78. These results are expected, as well as the pattern of comparative probabilities shown in the right-most figure. The lower-triangular part of the matrix shows higher probabilities that low concentration treatments outperform high concentration treatments.

\begin{figure}[h]
	\begin{center}
		\begin{tabular}{c}
			\includegraphics[height=.38\textwidth]{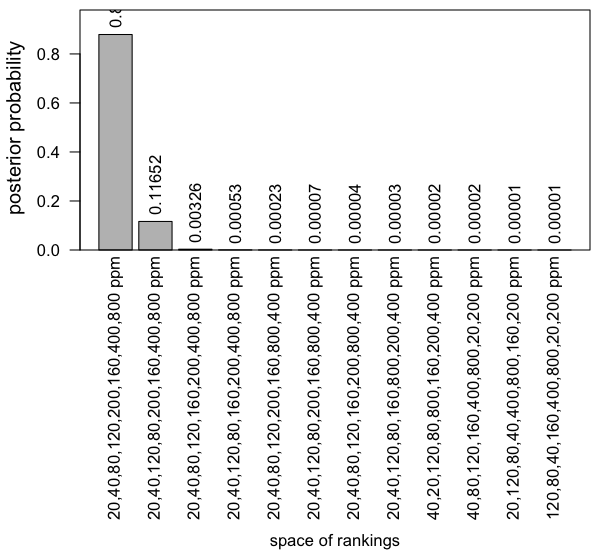}
		\end{tabular}
	\end{center}
	\vspace{-.2cm}
	\caption{Posterior distribution of rankings by minimum nitrogen concentration.}
	\label{fig:nitrogen_barplot}
\end{figure}

\label{sec:multivariate}
\subsection{Multivariate Data}

The task is now to identify the optimal treatment considering multiple objectives concurrently. In this case, the variables taken into analysis are plant net weight, the concentrations on the leaves of calcium, potassium, and sodium. Let us assume that we want to grow leafy crops commercialized for bone health. The joint evaluation criterion is now multivariate: we want plants with the highest net weight, the highest concentrations of calcium and potassium and, concurrently, the lowest concentration of sodium. For this example, we assume that the importance of dry weight doubles the importance of the concentrations of the minerals.

Following the idea introduced in \ref{sec:netweight}, settled theory in plant physiology suggests that, in an increasing gradient of treatments ordered by mineral concentrations, low levels correspond to deficit, intermediate levels to adequacy, and high levels to toxicity (e.g. \cite{Taiz_et_al_2015} and  \cite{Singh_and_Sharma_2014}). Because of this, it is expected that the yields produced under adequate levels outperform the yields under deficiency and toxicity. Then the optimal concentration for maximizing plant yield is such an adequate level that uses the least amount of nutrient. It is evident that though our approach does support that kind of analysis, is much more general as joint optimality for multiple features might come from ideal properties that are produced under any of those regimes designed thinking exclusively on yield. That is, using the BTME allows us to go beyond the triple (deficit, adequacy, toxicity).

\begin{figure*}[!t]
	\begin{center}
		\begin{tabular}{cccc}
			\includegraphics[height=.23\textwidth]{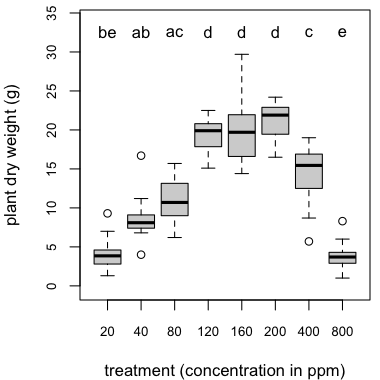} &
			\includegraphics[height=.23\textwidth]{fig/calcium_boxplot.png} &
			\includegraphics[height=.23\textwidth]{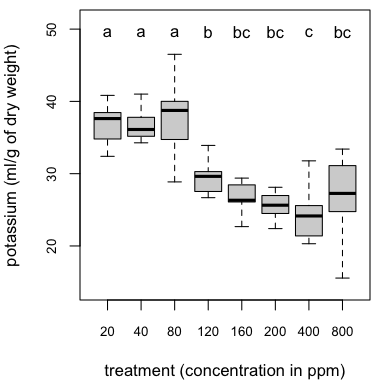} &
			\includegraphics[height=.23\textwidth]{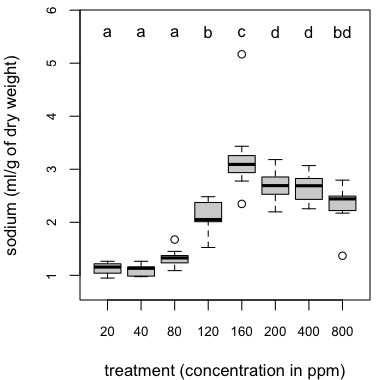} \\
			\includegraphics[height=.23\textwidth]{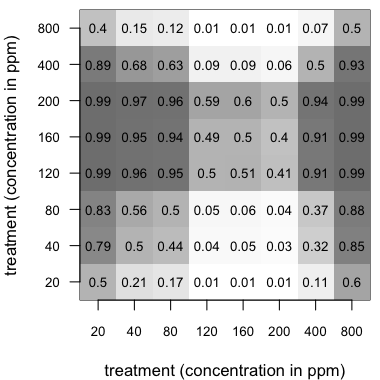} &
			\includegraphics[height=.23\textwidth]{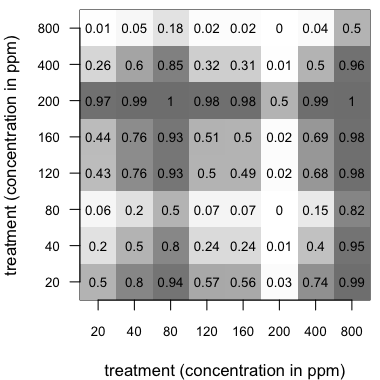} &
			\includegraphics[height=.23\textwidth]{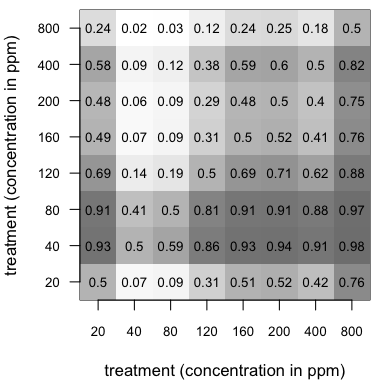} &
			\includegraphics[height=.23\textwidth]{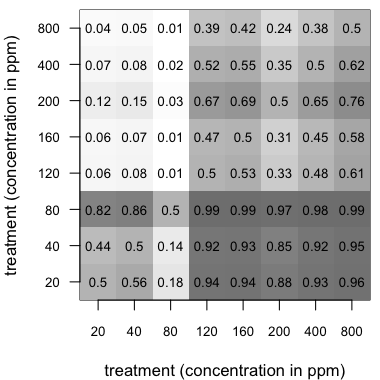}
		\end{tabular}
	\end{center}
	\vspace{-.2cm}
	\caption{First row: distributions of plant dry weight, and concentrations of calcium, potassium, and sodium, all by treatment. Second row: prevailing probabilities of producing better features: maximum dry weight, maximum calcium concentration, maximum potassium concentration, and minimum sodium concentration.}
	\label{fig:multivariate_boxplots_sigmoids}
\end{figure*}

The first row of figure \ref{fig:multivariate_boxplots_sigmoids} shows the distributions of plant dry weight, and concentrations of calcium, potassium, and sodium, all by treatment. When analyzed individually, central concentrations; 120, 160, and 200 ppm which happen to be non-significantly different from each other; clearly favor maximum plant dry weight and maximum calcium levels. Maximum levels of potassium are achieved by the treatments with lowest concentration, i.e. 20, 40 and 80 ppm, which clearly conflicts with the preferred treatments based on the first two criteria. Treatments 20, 40 and 80 ppm are also favored using sodium level as evaluation criterion.

The second row of figure \ref{fig:multivariate_boxplots_sigmoids} shows the prevailing probabilities of producing better features: maximum dry weight, maximum calcium concentration, maximum potassium concentration, and minimum sodium concentration. These results come directly from the joint estimation of global dominance indexes and weights and not from individual analysis. We highlight the virtue that estimation of by-feature and global treatment performances is being done cohesively in the sense of both being joint derivatives of the posterior inference process described in \ref{sec:inference} and computed using dominance indexes by feature as \ref{eq:dkj}.

Intuitively, the prevailing probability matrices show consistency with classical analysis. The one for plant dry weight show darker row lanes over 120, 160, and 200 ppm. For the calcium case the pattern of favorite treatments is less clear. However, the 200-ppm treatment exhibits the darkest row lane which is consistent. The darker lanes for potassium and sodium are 40, and 80 ppm; and 20, 40, and 80 ppm respectively, which is consistent with the treatments producing maximum potassium and minimum sodium, as desired.

\begin{figure}[h]
	\begin{center}
		\begin{tabular}{c}
			\includegraphics[height=.38\textwidth]{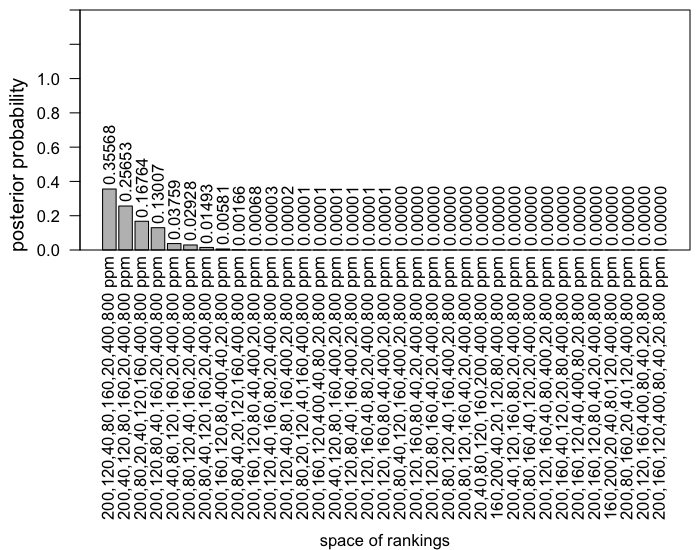}
		\end{tabular}
	\end{center}
	\vspace{-.2cm}
	\caption{Posterior distribution of rankings by the multivariate objective defined by maximum plant dry weight, maximum levels of calcium and potassium in the leaves, and minimum level of sodium in the leaves. The global ranking MAP estimate is (200, 120, 40, 80, 160, 20, 400, 800 ppm).}
	\label{fig:all_barplot}
\end{figure}

Finally, \ref{fig:all_barplot} shows the posterior distribution of rankings by the joint effect of four variables. The multivariate evaluation criteria is: maximum plant dry weight, maximum levels of calcium and potassium in the leaves, and minimum level of sodium in the leaves. The global ranking MAP estimate is (200, 120, 40, 80, 160, 20, 400, 800 ppm).

\section{Conclusion}

Although our method inherits the Bradley--Terry model assumptions, we made additional compromises (in \ref{sec:datastructure} and \ref{sec:likelihood}) inherent to our experiment-data setup to preserve the simplicity of the BTME. We discuss how they can be relaxed.

Regardless of the exact definition of the deviations, the BTME is not sensitive to the differences in the magnitudes of the original features. This happens due to the discrete nature of the comparisons which translate as counts. However, since the method integrates each feature equally into the likelihood, the sensitivity is translated at counts-level. For example, a treatment that performs very well under a feature that is not very relevant might be overrated. The way to deal with this is to incorporate important factors as tuning parameters, as exemplified in \ref{sec:multivariate}. In fact, these factors are more a desirable user input (capturing the notion of feature subjective importance) rather than a workaround to correct by the scale of the counts. Nevertheless, the study of the sensitivity of our outcomes to different normalization strategies is an open research avenue. This demands a careful design of a discrete normalization strategy, possibly leveraging the existing experience on continuous scaling. For example, \cite{Singh_and_Singh_2020}, \cite{Dalatu_and_Midi_2020}, and \cite{Singh_and_Singh_2022} have studied the sensitivity of popular machine learning techniques to specific data standardization choices.

The tractability of our method is only possible as we preserve the structural assumptions of independence over features, and independence over treatments. Naturally, other models can be derived by breaking these assumptions. The direct consequence is the inclusion of dependencies between the comparison random variables $Y_{jrs}$s. For example, it might be desirable to model explicitly the existing correlations between features. In our model, this is being compensated by imposing conditional independence over features given feature-level dominance indexes through the joint effect of $d$ and $w$. That is, by-feature ranks suffice for us to control by-feature treatment hierarchies. However, explicitly incorporating dependence may be appropriate. So far, dependence has been mostly studied as dependence among individuals, which is a feature we do not address. For example, \cite{Cattelan_2012} examines a family of paired comparison methods focused on dependence among individuals. The work of \cite{Altham_1978} and \cite{Altham_and_Hankin_2012} on the multivariate multiplicative binomial distribution provides a candidate building block to evolve the Bradley--Terry model by incorporating correlations at feature level.

Further directions of our model include applying it in predictive analysis. For example, ideal expected responses might come from treatments that were never conducted. Thus, we can use observed-treatment outcomes to predict the performance of unseen treatments, allowing for the possibility that any of those is the actual optimal. This problem can be seen as the missing data (\cite{Rubin_1976}) extension of our model in (\ref{eq:likelihood}). That is, to consider a sampling distribution for complete data $p(\boldsymbol{y}_{\text{obs}},\boldsymbol{y}_{\text{mis}} \mid \boldsymbol{n},\boldsymbol{\theta})$ ($y$s defined as in \ref{eq:Y}) where the obs-subscripted $\boldsymbol{y}$ represents the observable $y_{jrs}$s, and the mis-subscripted $\boldsymbol{y}$ the $y_{jrs}$s encoding the comparisons of deviations with respect to unobserved $r$ or $s$ treatments. We are unable to propose yet a specific form for the dependence of $\boldsymbol{y}_{\text{mis}}$ on $\boldsymbol{y}_{\text{obs}}$ and the best strategy to sample unseen treatments. These characterizations require a refined analysis and merit a separate investigation.

A whole family of applications arises when fitting our model controlling by non-individual-level variables, which play a different role than responses recorded at individual-level. For example, suppose (ignoring any practical implementation for now) we execute of our model daily for the plant phenotyping experiments from section \ref{sec:application}. Every day, data of environmental conditions is recorded (apart from plant-level features) and thus observing the rankings changing dynamically is feasible. Specifically, computing the MAP estimate (as in equation \ref{eq:est}) from daily data allows us to visualize the reallocation of treatments in the ranking over time. Systematic treatment-position swapping informs about the best treatment for different environmental conditions.

Regardless of the considerations and limitations discussed, we have reported promising results from using our model in experiment evaluation. It can achieve better understanding of general experiment data and offers a new way to visualize comparative results. Most importantly, it reshapes classic treatment evaluation by leveraging a distribution on the space of treatment rankings, which arguably constitutes a more natural solution for the question {\em which treatment is the optimal?}.

Finally, we aim to contemporize the discussion and application of ranking inference methods on trials-based fields, an area that has been largely overlooked in contemporary AI. Our method's characteristics make it particularly suitable for extensions in RL. Arguably, the concept of ranking-based RL is not easy to conceive without the presented framework. Sampling treatments based on their rank and using feedback from a multivariate state space to dynamically update the ranking distribution offers a direction for implementing Thompson sampling exploration.

\section*{Acknowledgment}

This research is supported by the National Research Foundation, Prime Minister's Office, Singapore, under its Campus for Research Excellence and Technological Enterprise (CREATE) and the Singapore-HUJ Alliance for Research and Enterprise (SHARE), program REQ0246254 (NRF2020-THE003-0008). We gratefully acknowledge assistance from our collaborators Netatech Pte Ltd, Singapore, Mr. David Tan, and Dr. Magdiel Inggrid Setyawati.

\ifCLASSOPTIONcaptionsoff
  \newpage
\fi

\bibliographystyle{IEEEtran}
\bibliography{Bibliography}

\end{document}